\documentclass[12pt]{article}
\usepackage[utf8]{inputenc}
\usepackage[total={16cm,24cm},centering]{geometry}
\usepackage{parskip}
\usepackage{multicol}
\usepackage{url}
\usepackage{cite}
\usepackage{float, graphicx, caption, color, amsmath, parskip, fancyhdr}
\usepackage{subcaption}
\graphicspath{{img/}}

\title{\textbf{OPTIMIZATION OF THE ANALYSIS OF
DATA OF VITAL EVENTS
USING THREADS}}

\author{Coasaca Callacondo Rubi melania \\\textcolor{blue}{rcoasacac@est.unap.edu.pe} \\Chata Iscarra Grylia Yaneth \\ \textcolor{blue}{gchatai@est.unap.edu.pe} \\ Fred Torres-Cruz \\\textcolor{blue}{ftorres@unap.edu.pe}}
\date{28 November 2022}

\begin{document}
\maketitle
\thispagestyle{fancy}
\section*{Abstra}
\textbf{Objective:} Optimize the time of data analysis of Vital Events (births,
deaths and marriages) using Threads.

\textbf{Methodology:} A code was created in python without threads and another with threads, after
He performed 5 tests with a single attribute and with 1,2,3,4,5,6,7,8 and 9 attributes this was done
in both codes with the same amount of data, to know if the python code with
threads is optimal when parsing Vital Facts data.
 
\textbf{Results:} The python code with Threads turned out to be the most optimal since it optimized
the compilation time of the 5 tests with 1 attribute by 16
attributes was obtained as a result that the more attributes you group, the more effective the
use of threads.

\textbf{Conclusion:} Python code with threads is more optimal than code without threads
Therefore, it is concluded that the implementation of threads is recommended in the analysis of
data in similar works.

\begin{multicols}{2}

\section*{Introduction:} 
Peru generates millions of data records daily, one of those records are the Vital Facts (births, deaths and marriages), this record is increasing day by day due to the fact that record events are generated daily. As there is such a large amount of data, the analysis of said data takes time and as a solution to this problem, what is intended in this study is to optimize the analysis of 500,000 data from the Vital Events registry, implementing a code in python where the data analysis will be optimized through parallel computing, implementing threads. This optimization will be demonstrated by the read time and compile count of 9 attributes.

\section*{Methodology:}

\textbf{Parallel computing:} Computing parallel divides a big problem into several parts small, separate, sometimes similar can be executed simultaneously by multiple processors that communicate through shared memory, the results are finally they execute as if the work had been done in a single general algorithm.

\textbf{Threads:} Threads are multiple threads running simultaneously in parallel and performing different tasks in a single program, \cite{ibanez2022paralelizacion} Memory-sharing threads will be created to speed up the process of reading and counting the 500000 data with 9 attributes.

\textbf{Data:}
The data corresponds to the National Open Data Platform, Governance Vital Facts option, which can be viewed at the following link:\textcolor{blue}{\url{bit.ly/3udD9Wh}} This information is provided by the National Institute of Statistics and Information ethics - INEI, source-RENIEC.

\textbf{Data description:}
We will work with 500,000 vital event data that are in CSV format, containing 9 attributes. The last time this data was updated was in September 2022. The data contains registrations of births, deaths, and marriages registered in civil status registry offices nationwide, broken down by year and month of registration, sex, vital event, type of registration, department, province and district where the registration was made.

\textbf{Data Dictionary:}

\begin{figure}[H]
    \centering
    \includegraphics[width=7.85cm]{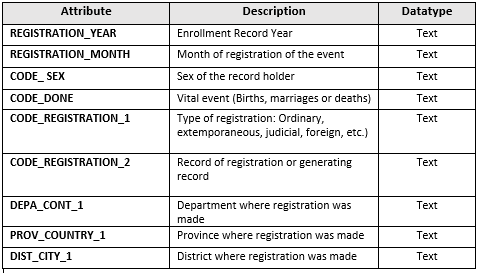}
\end{figure}

\textbf{Implementation Logic:} A code will be made in python language where parallel computation will be applied implementing threads  \cite{keba2021aplicacion}\\

\vspace{ 0.03cm }
- Load the data of Vital Facts that are in CSV format, the code is in python language but it is executed in google colab.\\
 \vspace{ 0.03cm }
- The code starts reading and counting the data until it finishes all the traversal of the requested attribute.\\
  \vspace{ 0.03cm }
- Performs a reading and counting of 500,000 data, with 1, 2 or up to 9 attributes that are indicated to the code.\\
 \vspace{ 0.03cm }
 - The reading and counting of the 500000 data is done in parallel, shortening the execution time through the implementation of threads.
 
\textbf{Implementation:} The following libraries, functions and coherence were used to make the python codes with threads and without threads. -the pandas library is used for the export and processing of the 500000 data.\\
\vspace{ 0.03cm }\\
-The time library is in charge of controlling the execution time of the data.\\
\vspace{ 0.03cm }\\
-The threading library allows code to execute multiple operations simultaneously.\\
\vspace{ 0.03cm }\\
-The Numpy library is used to create matrices faster and store a large amount of information.\cite{andrango2022implementacion}\\
\vspace{ 0.03cm }\\
-The data is imported, we continue with the extraction of the data attributes; and select the attributes to work with.\\
\vspace{ 0.03cm }\\
-The timer starts once the attributes have been assigned to it.\\
\vspace{ 0.03cm }\\
-The function count data is in charge of counting all the data that belongs to the requested attribute.\\
\vspace{ 0.03cm }\\
-It begins with the first attribute that was assigned and its respective reading and count.\\
\vspace{ 0.03cm }\\
-Successively, it is passed from attribute to attribute with its respective count according to the requested attribute.\\
\vspace{ 0.05cm }\\
-Prints the amount of data and the time it took to count and read it.\\
\vspace{ 0.05cm }\\
\textbf{Pseudocode:} Pseudocode is one of the most widely used methods for designing algorithms, and it is independent of the programming language that is going to be used.\cite{lopez2022experiencias}

\textbf{Pseudocode: \cite{avalos1}}
\begin{figure}[H]
    \centering
    \includegraphics[width=8cm]{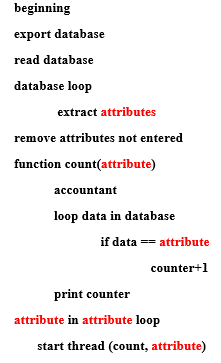}
\end{figure}

\textbf{Libraries and threadless code:} 
\begin{figure}[H]
    \centering
    \includegraphics[width=8cm]{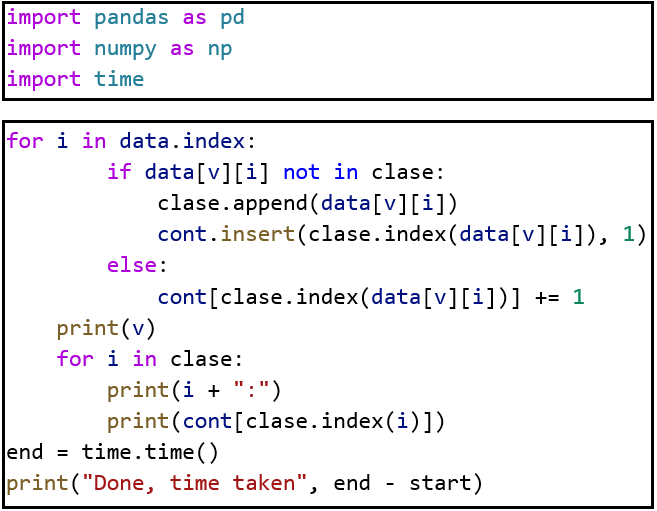}
\end{figure}

\textbf{libraries and code with threads}:

\begin{figure}[H]
    \centering
    \includegraphics[width=8cm]{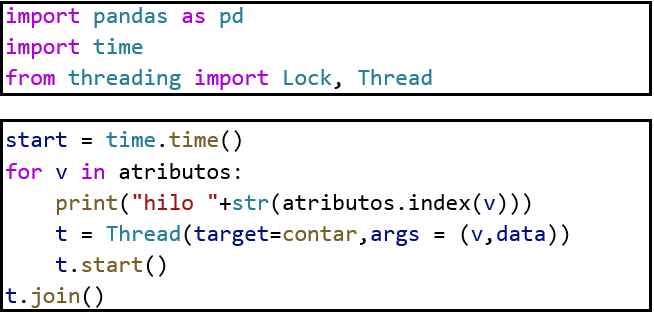}
\end{figure}

The python code is shown where it performs a reading and a count of the data according to the requested attributes, then only the part of lines of code where the threads are implemented.

\section*{Results:}
In both codes the 500000 were compiled data of vital events first with 1 attribute, 2 attributes,..., until completing the 9 attributes, likewise 5 tests were carried out with 5 independent attributes (year, month of registration, sex, event and registration 1) .

\textbf{Time difference with 9 attributes:}

\begin{figure}[H]
    \centering
    \includegraphics[width=8cm]{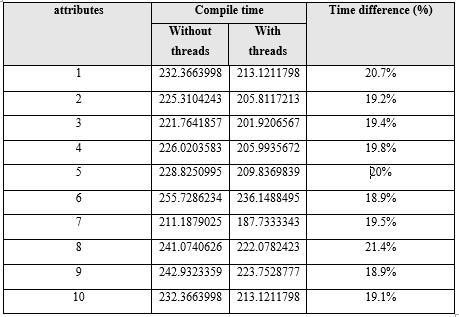}
\end{figure}

\textbf{Graphic:}
\begin{figure}[H]
    \centering
    \includegraphics[width=8cm]{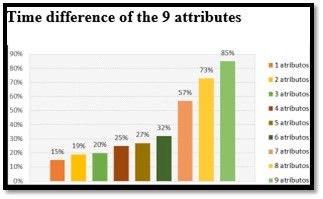}
\end{figure}

\textbf{Interpretation:}
In the Bar Graph it is visualized that the difference that exists between attribute 1 and attribute 9 is wide, it is also observed that the bars are in ascending order, this leads us to say that the more attributes are compiled, the more effective the use of threads.

\textbf{mean test:}
\begin{figure}[H]
    \centering
    \includegraphics[width=8cm]{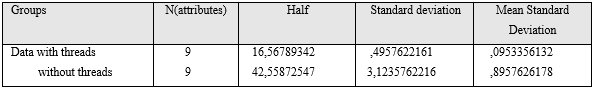}
\end{figure}

\textbf{Interpretation:}

According to the table, it can be seen that in the data analyzed without optimization we have an average of 42.55872547 seconds while with optimization we have an average of 16.56789342 seconds, which indicates that it is better to use optimization through threads.
Therefore, we say that we can obtain response times and the percentage of improvement according to the number of attributes analyzed, resulting in a percentage of between 60\% to 65\% improvement in each test, as can be seen in the following table.

\begin{figure}[H]
    \centering
    \includegraphics[width=8cm]{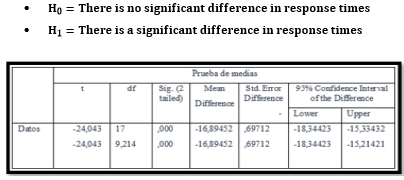}
\end{figure}

\textbf{Interpretation:} 
Applying the means test on the data obtained with an Alpha equal to 0.05, we reject the null hypothesis, which means that if there is a
significant difference in the response when comparing both in the year
and registration of a total of 500000 data.

\textbf{Time difference of 5 tests:}

\begin{figure}[H]
    \centering
    \includegraphics[width=8cm]{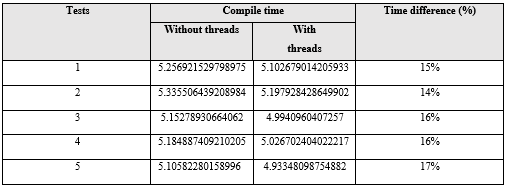}
\end{figure}

\textbf{Graphic:}

\begin{figure}[H]
    \centering
    \includegraphics[width=8cm]{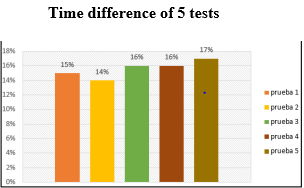}
\end{figure}

\textbf{Interpretation:}
The bar graph shows that the difference between test 1 and test 5 is almost similar, this leads us to say that by individually compiling each attribute, a 16\% improvement can be seen when using threads.

\textbf{average:}
To do this we will use the following formula:

\begin{figure}[H]
    \centering
    \includegraphics[width=5cm]{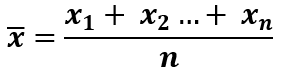}
\end{figure}

\begin{figure}[H]
    \centering
    \includegraphics[width=8cm]{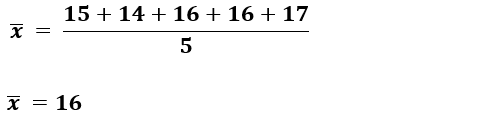}
\end{figure}

\textbf{Interpretation:} The average of the differences of the 5 tests with independent attributes between the python code with threads and the python code without threads is 16\%.

\section*{Conclution:} A noticeable difference is seen between the difference of the 9 attributes as well as in the 5 tests carried out on individual attributes using threads. Therefore, it is concluded that using threads is effective for similar cases where data analysis is required to be more optimal.
\section*{Observations on the data:}

\textbf{Graphic attribute Sex:}

\begin{figure}[H]
    \centering
    \includegraphics[width=6cm]{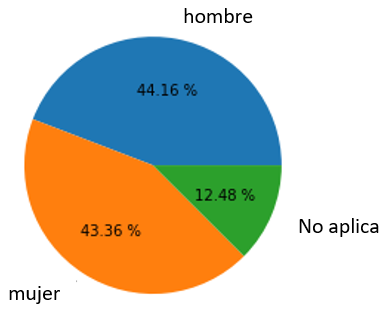}
\end{figure}

\textbf{Interpretation:}
The pie chart shows us that the holder of the document was 44.16\% male, 43.36\% female, and 12.48\% not applicable. in the registry of vital events with 500,000 data. From which it can be deduced that it was men and women who were the most holders of the minutes while it did not apply, it was very low. 

\textbf{Graph attribute Done:}

\begin{figure}[H]
    \centering
    \includegraphics[width=6cm]{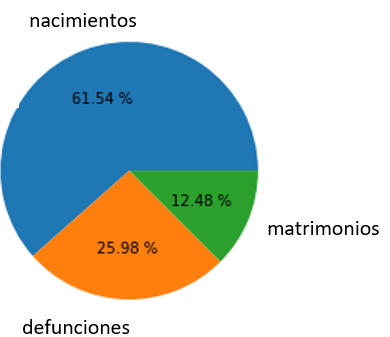}
\end{figure}

\textbf{Interpretation:}
The pie chart shows us that there were 61.54\% births, 25.98\% deaths and 12.48\% marriages. in the registry of vital events with 500,000 data. From which it can be deduced that there were more births than deaths and marriages and it is also visualized that the percentage of marriages is half of the deaths.

\section*{Discussion:}

It was proposed to reduce the processing time of large amounts of data in a case of three measures of semantic similarity in the field of biology using parallel processing, as a result a reduction in time was achieved that gradually increases with the increase in data.\cite{Almasoud2019}

Visual information data processing is important in homes when making a comparison of which language adjusts to the processing conditions, as a result it was obtained that the code is more effective in Python processing images in parallel using threads since it receives the same image but processing it on different threads.\cite{Bustamante2022}

In general, data processing also usually includes queries, algorithms for query processing of context-free routes, this can include an endless number of symbols which reflects execution time. It was evaluated through tests in Go and Python, a biology database was used and as a result, performance gains were demonstrated when using threads and not only working with a single process.\cite{Medeiros2022}

Projecting billions and millions of longitude and latitude location data onto hundreds of thousands of highly irregular population census block polygons is computationally challenging and can be implemented in any scripting language such as Python. the tests were divided into simple and fast as a result it is appreciated that the fast approach exploits the optimizations through threads, achieving quick integration of location and demographic data.\cite{Kepner2020}

If it is proposed to implement and evaluate two error propagation metrics first error propagation through structured data second location corruption fraction both software and hardware, both metrics were evaluated by running single threads and multiple threads, which demonstrated that using multithreading is more effective detecting both metrics.\cite{Ozturk202218691}

It is observed that the efficiency of the journey optimizes and speeds up the processing time of large amounts of data in the broader field of analysis through open source projects and industry projects, using parallel processing and as a result a potential reduction was achieved. of time that is beneficially increased by the increase in data mining.\cite{heseding2022tooling}

It has been carried out in order to process the large data on the Markov chain Monte Carlo Algorithm, it was possible to observe the speed of time was considerable and satisfactory at the moment of parallelization processing through the implementation of threads that optimize and is the best handling of reads at compile time.\cite{li2021multithreaded}

an explosion of software tools have been created to process these data files in order to facilitate this, a library was produced from the original SAMtools implementation, with a focus on performance and robustness performance has improved considerably, with a BAM read and write loop that runs 5 times faster and a conversion from BAM to SAM 13 times faster and at the time of processing the large data it was possible to observe the speed of time was satisfactory at the time of parallelization processing through the threads implementation that optimizes and is the best when compiling.\cite{bonfield2021htslib}

Especially in data development parsing algorithms While true when processing data and compiling code with threads they provide powerful tools to speed up the development process, on the other hand, by specializing the interpreter for a given script. Loops in the specialized code can then be parallelized to further improve performance time.\cite{neth2019automatic}

A code was implemented in order to optimize the time in parallel with threads when processing a large amount of data that have made possible important advances in computer science thanks to the computing power and the consequent impossibility of analyzing and understanding the results with the current techniques. The Neurolytics implementation is primarily focused on simulation neuroscience to help scientists structure and understand the results of their experiments in a fast and efficient way. \cite{planas2018accelerating}

\bibliographystyle{plain}
\bibliography{bibliografia.bib}
\end{multicols}
\end{document}